\documentclass[3p,times]{elsarticle}
\usepackage{ecrc}

\volume{00}

\firstpage{1}

\journalname{Annals of Physics}

\runauth{S. Longhi}

\usepackage{amssymb}
\usepackage[figuresright]{rotating}

\begin{document}

\begin{frontmatter}



\title{Phase transitions in Wick-rotated $\mathcal{PT}$-symmetric optics}


\author{Stefano Longhi}
\address{Dipartimento di Fisica, Politecnico di Milano and Istituto di Fotonica e Nanotecnologie del Consiglio Nazionale delle Ricerche, Piazza L. da Vinci 32, I-20133 Milano, Italy\\ Tel/Fax: 0039 022399 6156/6126, email: longhi@fisi.polimi.it}

\begin{abstract}
$\mathcal{PT}$-symmetric models with a Wick rotation of time ($ t \rightarrow \pm i t$) show  spectral phase transitions that are similar to those of dissipative systems driven out of equilibrium. Optics can provide an accessible test bed to explore spectral phase transitions of Wick-rotated $\mathcal{PT}$-symmetric models. This is shown by considering the transverse dynamics of laser light in optical cavities with variable reflectivity and tilted mirrors.
Two specific examples are discussed: the optical analogue of the hydrodynamic Squire model of vorticity, and the Wick-rotated $\mathcal{PT}$-symmetric nonlinear dimer model. In the latter case the spectral phase transition is associated with the universal phase locking-unlocking transition in Adler's theory of  coupled oscillators.
\end{abstract}


\end{frontmatter}

\section{Introduction}

In 1998, Bender and Boettcher  showed that a wide class of non-Hermitian
Hamiltonians $\hat{H}$ can possess entirely real spectra as long as they respect parity-time ($\mathcal{PT}$) symmetry \cite{Bender0,Bender1}.
While the implications of $\mathcal{PT}$ symmetry in theoretical physics are still a matter
of debate \cite{debate}, classical systems such as optical \cite{opt1,opt2,opt3,opt3bis,opt4,opt5,opt6} and electronic \cite{el0} systems provide an accessible test bed where the  $\mathcal{PT}$ symmetry notion can be explored. $\mathcal{PT}$-symmetric Hamiltonians 
 show a sharp spectral transition when a control parameter is varied, with the appearance of pairs of complex conjugate energies in the broken $\mathcal{PT}$ phase. 
The phase transition is associated with the appearance of exceptional points \cite{Klaiman} or spectral singularities \cite{Mosta,Longhi}. Similar transitions are found in pseudo-Hermitian Hamiltonians \cite{pseudo1}, pseudo-$\mathcal{PT}$-symmetric driven Hamiltonians \cite{pseudo2}, Wick-rotated $\mathcal{PT}$ symmetric Hamiltonians \cite{squire0}, and in Liouvillean operators in the Lindblad form \cite{referee}.  In particular, in Ref.\cite{referee} it was shown that a combination of unitary and antiunitary symmetry of quantum Liouvilleans associated to certain open quantum systems implies a dihedral symmetry of the complex Liouvillean spectrum. A different (and not necessarily dihedral) symmetry of the spectrum is found in Hamiltonian systems after application of Wick rotation \cite{Wick}, which consists  in rotating the time axis by $\pm \pi/2$ in complex plane, i.e. to the transformation $t \rightarrow \pm it$. In particular, Wick rotation of $\mathcal{PT}$ symmetric Hamiltonians generates the same spectral transition in complex plane, but rotated by $\pm \pi/2$.  Complexification of time and corresponding rotation of the spectrum deeply changes the physical signatures of the symmetry breaking. Physical systems described by Wick-rotated $\mathcal{PT}$-symmetric models are found in hydrodynamics \cite{squire0,squire1} and in certain gauge field theories \cite{Yang}.
A paradigmatic example is the Squire model, which was introduced in hydrodynamics to describes the normal vorticity of a plane Couette flow \cite{squire0,squire1,squire2} and to explain large transient growths of perturbations in spite of the linear stability of the underlying flow \cite{squire1,squire3,modelocking}.\\
In this work we show that spectral phase transitions in Wick-rotated $\mathcal{PT}$-symmetric Hamiltonians behave like phase transitions in dissipative systems driven out of equilibrium, and that optics can provide an accessible test bed where such spectral phase transitions and their physical signatures can be explored. 
We consider transverse laser dynamics in optical resonators with variable reflectivity mirrors \cite{Siegman} and discuss, as examples, the optical realization of the hydrodynamic Squire model \cite{squire1,squire2} and the Wick-rotated nonlinear $\mathcal{PT}$-symmetric dimer model \cite{dimer1,dimer2}.

\section{Optical Resonator Model of Wick-rotated $\mathcal{PT}$-symmetric systems}
A dynamical system that realizes a Wick-rotated $\mathcal{PT}$-symmetric model is described quite generally by the following equation for an order parameter $\psi(x,t)$
\begin{equation}
\partial_t \psi= -\hat{H} \psi- |\psi|^2 \psi 
\end{equation}
where $\hat{H}$ is the $\mathcal{PT}$-symmetric operator defined by
\begin{equation}
\hat{H}=-\partial^2_x+V(x)-g_0.
\end{equation}
In Eq.(2), $V(-x)=V^*(x)$ is the $\mathcal{PT}$-symmetric complex potential, whereas $g_0$ is a constant real parameter that just provides a shift of the real part of the energies of $\hat{H}$. In Eq.(1), a cubic nonlinear term is added to limit the growth of unstable modes of $\hat{H}$.  
In optics, a possible realization of Eq.(1) is provided by transverse laser dynamics in an optical resonator with variable reflectivity and aspherical mirrors. Let us consider the optical cavity shown in Fig.1 with a one spatial transverse coordinate $X$. The resonator comprises two end mirrors in a nearly self-imaging configuration \cite{Weiss}, one totally-reflective flat mirror (mirror 2) and the other one a variable-reflectivity and aspherical mirror (mirror 1). The reflectivity of mirror 1 is given by $r(X)=\sqrt{R(X)} \exp[i \Delta(X)]$, where $R(X)$ and $\Delta(X)$ are the transversely-varying power reflectance and phase shift introduced by the aspherical surface of the mirror. The gain medium, placed close to mirror 1, is assumed to have a fast polarization and population relaxation rates (class-A laser \cite{Weiss,Arecchi}; e.g. He-Ne, Ar$^+$, Kr$^+$ or dye lasers). A set of two focusing lens of focal length $f$  and a Gaussian aperture, placed in the focal plane of the lenses, provide spectral filtering of the optical field at the plane $\gamma$ in the cavity \cite{Weiss}; see Fig.1. Indicating by $t(X)=\exp(-X^2/w_a^2)$ the spectral transmission of the Gaussian aperture of size $w_a$ and neglecting diffraction (propagative) effects in the gain medium, assuming single-longitudinal mode oscillation the evolution of the electric field envelope $E(X,T)$ at plane $\gamma$ in the cavity is governed by the following equation (see Appendix A for details)
 \begin{figure}
\includegraphics[scale=0.3]{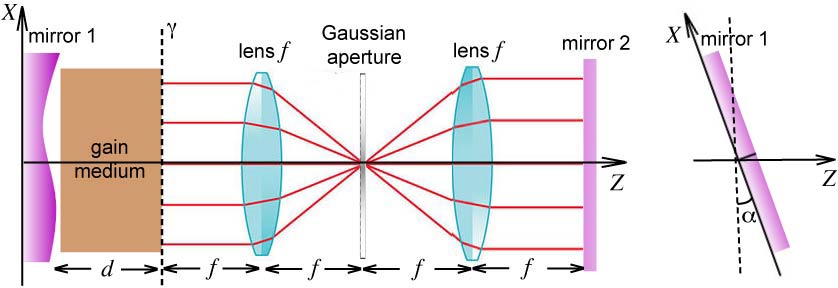}
\caption{(Color online) Schematic of a self-imaging optical resonator. Mirror 1 is an aspherical and variable-reflectivity mirror with reflection coefficient $r(X)=\sqrt{R(X)} \exp[i \Delta (X)]$, whereas mirror 2 is a flat and fully reflective mirror. The right panel shows the case of a flat tilted mirror.}
\end{figure}
\begin{equation}
T_R \partial_T  E=\mathcal{D} \partial^2_X E -V(X) E+gE
\end{equation}  
where $T_R$ is the photon round-trip time in the cavity, $g$ is the double-pass saturated gain in the active medium, $\mathcal{D}=  f^2 \lambda^2/(2 \pi^2 w_a^2)$ is the spatial spectral filtering parameter, $\lambda$ is the laser wavelength, and 
\begin{equation}
V(X)=-{\rm ln} [ r(X)] = -{\rm ln} \sqrt{R(X)}-i \Delta (X).
\end{equation}
For class-A lasers the saturated gain $g$ is simply given by \cite{Weiss} $g=g_0/(1+ \chi |E|^2)$, where $g_0$ is the small-signal (unsaturated) gain parameter that depends on the pump rate and $\chi$ is inversely proportional to the saturation intensity of the laser transition. Close to laser threshold, the Lamb expansion holds, i.e. $g \simeq g_0 (1- \chi |E|^2)$, which provides the simplest model of gain saturation  \cite{Arecchi}. After the introduction of normalized space and time variables $t=T/T_R$, $x=X/L$ with characteristic spatial length $L=\sqrt{\mathcal{D}}= \lambda f /(\sqrt{2} \pi w_a)$ and after setting $\psi= \sqrt{g_0 \chi} E$,  the field evolution inside the optical resonator can be set in the canonical form Eq.(1), where the real and imaginary parts of the potential $V(x)$ are determined by the spectral reflectance and aspherical surface shape of the mirror 1 according to Eq.(4). $\mathcal{PT}$ symmetry of $\hat{H}$ requires that $R(x)$ and $\Delta(x)$ be even and odd functions of $x$, respectively. An experimentally simple and interesting case is that of a tilted mirror \cite{Staliun}  of variable reflectance $R(X)=R(-X)$ (right panel in Fig.1). Indicating by $\alpha$ the tilting angle, one has $\Delta(X)= -(2 \pi / \lambda) \alpha X$ and thus 
\begin{equation}
V(x)=-{\rm ln} \sqrt{R (xL)}+i \gamma x
\end{equation} 
where $\gamma=(2 \pi / \lambda) \alpha L$. In the following we will present two significant examples of spectral phase transitions of Eq.(1) with an optical potential of the form of Eq.(5): the optical analogue of the hydrodynamic Squire model \cite{squire2,squire3} and the Wick-rotated $\mathcal{PT}$-symmetric nonlinear dimer model \cite{dimer1,dimer2}. \par
 \begin{figure*}
\includegraphics[scale=0.24]{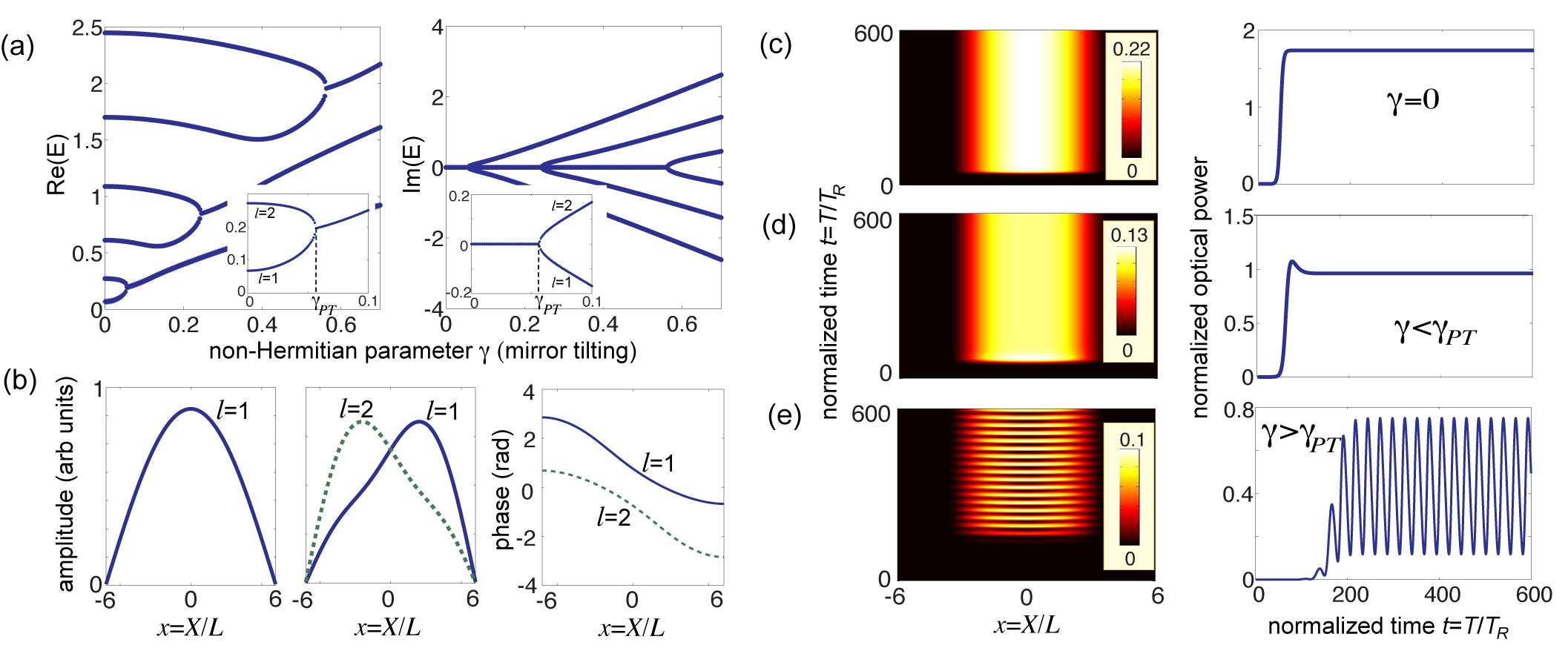}
\caption{(Color online) Optical Squire model. (a) Energy spectrum (real and imaginary parts) versus $\gamma$ of the $\mathcal{PT}$-symmetric Squire operator $\hat{H}=-\partial^2_x+i \gamma x$ in the interval $(-u,u)$ for $u=6$. $\mathcal{PT}$ symmetry breaking is attained at $\gamma_{PT} \simeq 0.056$. The insets show  the details of lowest-order eigenvalue coalescence leading to symmetry breaking transition. (b) Profile of the lowest-order resonator mode ($l=1$) in the Hermitian case $\gamma=0$ (left panel), and of the two threshold-degenerate modes ($l=1,2$) at $\gamma=0.07$ (central and right panels). (c-e) Numerically-computed evolution of normalized transverse light intensity $|\psi(x,t)|^2$ (left panels) and optical power $P(t)= \int dx |\psi(x,t)|^2$ (right panels), showing laser switch on from an initially small-amplitude random noise. The unsaturated gain parameter is  $g_0=0.25$, whereas the normalized tilting angle is $\gamma=0$ in (c), $\gamma=0.04$ in (d), and $\gamma=0.07$ in (e). Physical units are $L \sim 57 \; \mu$m and $T_R \simeq 4$ ns; tilting angle corresponding to $\gamma_{PT}$ is $\alpha_{PT} \simeq 0.1 \;$mrad.}
\end{figure*}
\section{ Optical Squire model.} Let us consider the case where the reflectance $R(X)$ of mirror 1 is a top-hat function, i.e. $R=1$ for $|X|<a$ and $R=0$ at $|X|>a$, where $2a$ is the finite mirror aperture. The real part of the potential (4), ${\rm Re}(V)=-{\rm ln} \sqrt{R(xL)}$, describes an infinite potential well, i.e. it vanishes for $|x|<u$ and becomes infinite at the boundaries $x= \pm u$, where $u=a/L$. In this limit $V(x)=i \gamma x$ and Eq.(1) should be integrated in the interval $(-u,u)$ with the boundary conditions $\psi(-u)=\psi(u)=0$: this is precisely the Squire equation  that describes the normal vorticity of a plane Couette flow with linear velocity profile \cite{squire0,squire1,squire2}. A typical example of numerically-computed spectrum of the Squire operator as a function of mirror tilting $\gamma$ is shown in Fig.2(a) for $u=6$. Note that a transition from an entirely real spectrum to pairs of complex conjugate energies occurs at the $\mathcal{PT}$ symmetry breaking point $\gamma_{PT} \simeq 0.056 $, where the two lowest real energies coalesce yielding a pair of complex conjugate energies  [see the insets in Fig.2(a)].  
For $\gamma<\gamma_T$, all eigenvalues of $\hat{H}$ are real and positive; in particular, at $\gamma=0$ the  energies are those of an infinite potential well and are given by $E_l= (l \pi / 2 u)^2$, where $l=1,2,3,...$ is the transverse mode index. As
 the gain $g_0$ is increased, the most unstable mode that reaches threshold is the mode with index $l=1$, which is depicted in Fig.2(b), left panel.  
 Above threshold a steady-state single transverse mode operation is observed owing to gain saturation; see Fig.2(c). A similar scenario is observed for a non vanishing value of $\gamma$ below $\gamma_T$; see Fig.2(d). Conversely, above  the $\mathcal{PT}$ symmetry breaking point, $\gamma> \gamma_{PT}$, there is a couple of  transverse modes, with the same gain threshold but different oscillation frequencies that can compete. An example of the profiles of such modes is shown in the central and right panels of Fig.2(b). A weakly nonlinear analysis of Eq.(1) above threshold shows that both modes oscillate, because self-saturation prevails over cross-saturation  of the gain. The technical details are given in Appendix B. Correspondingly, as $\gamma$ is increased above $\gamma_{PT}$ the laser ceases to emit a continuous-wave power, and an oscillatory behavior is observed as a result of mode beating and mode non-orthogonality; see Fig.2(e). The transition from a stationary to an oscillatory laser emission is thus the clear signature of the $\mathcal{PT}$ symmetry breaking transition in the Wick-rotated Hamiltonian. To get an idea of physical parameters corresponding to the symmetry breaking transition, let us consider as an example an optical cavity for a He-Ne laser ($\lambda=633$ nm) and let us assume a Gaussian aperture of size $w_g=250 \; \mu$m, a focal length $f=10$ cm, and a total cavity length $\mathcal{L}=4f+d=60$ cm. The characteristic spatial and temporal scales are $L \sim 57 \; \mu$m and $T_R \simeq 2 \mathcal{L}/c \simeq 4$ ns, so that the transverse aperture of mirror 1 is  $ 2a=2uL\simeq 684 \; \mu$m, the symmetry breaking transition, from a stationary to an oscillatory laser output power, occurs at the tilting angle $\alpha_{PT}=\gamma_{PT} \lambda /(2 \pi L) \simeq 0.1$ mrad; finally, the temporal period of power oscillations in Fig.2(e) is $\sim 108 $ ns. 
  \begin{figure*}
\includegraphics[scale=0.28]{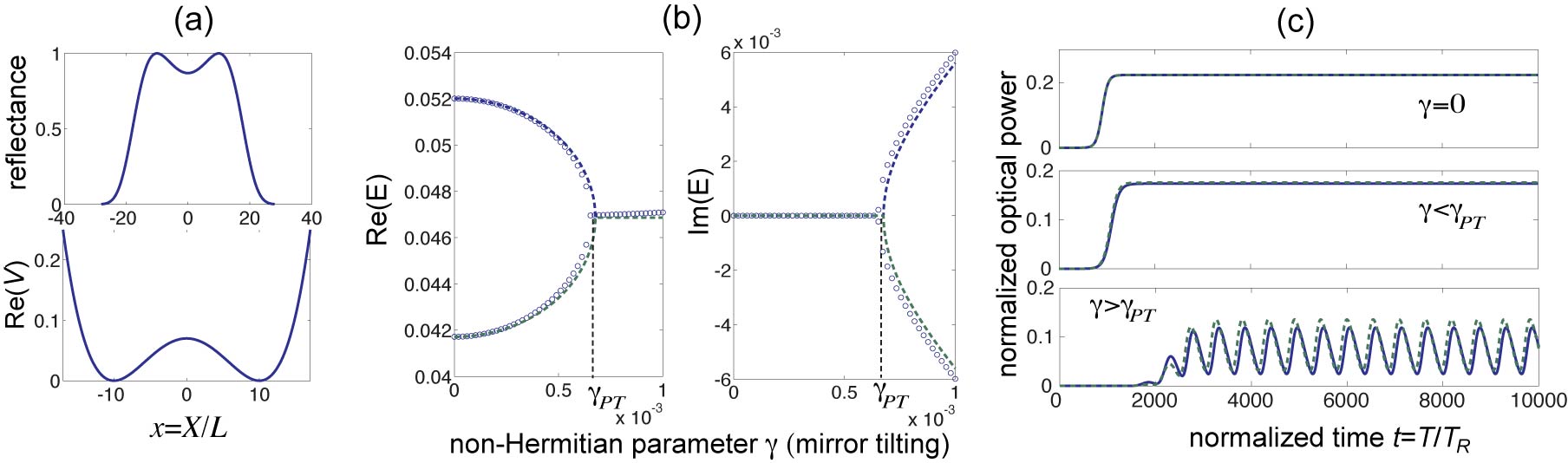}
\caption{(Color online) Wick-rotated $\mathcal{PT}$ symmetric dimer model. (a) Mirror reflectance $R(x)$ (upper panel), corresponding to the double well potential ${\rm Re}(V)=\beta(x^2-x_0^2)^2$ (lower panel) for $\beta=7 \times 10^{-6}$ and $x_0=10$. (b) Energy spectrum of $\hat{H}$ (circles) versus $\gamma$. 
 $\mathcal{PT}$ symmetry breaking is attained at $\gamma_{PT} \simeq 0.00065$. The dashed curves show the energy spectrum as predicted by the reduced model [Eqs.(6) and (7)].  (c) Numerically-computed evolution of normalized optical power $P(t)= \int dx |\psi(x,t)|^2$ (solid curves), showing laser switch on from an initially small-amplitude random noise. The unsaturated gain parameter is  $g_0=0.05$, whereas the normalized tilting angle is $\gamma=0$ in the upper plot, $\gamma=0.0005$ in the middle plot, and $\gamma=0.0007$ in the lower panel. The dashed curves are obtained from a numerical solution of the reduced model assuming $\kappa \simeq 0.00515$, $q \simeq 0.0469$, $\rho \simeq 0.074$ and $\sigma \simeq 7.72 \gamma$. Physical units are $L \sim 57 \; \mu$m and $T_R \simeq 4$ ns; tilting angle corresponding to $\gamma_{PT}$ is $\alpha_{PT} \simeq 1.2 \; \mu$rad.}
\end{figure*}

\section{Wick-rotated $\mathcal{PT}$-symmetric dimer model.}
As a second example, we consider the Wick-rotated nonlinear $\mathcal{PT}$-symmetric dimeric model, which is obtained by assuming in Eq.(5) a double well shape for the real part of $V(x)$. Without Wick rotation this model was studied in Refs.\cite{dimer1,dimer2}, where  coexistence of periodic and blow up (unbounded) solutions  was predicted in the broken $\mathcal{PT}$ phase.  An example of two-hamped mirror reflectance $R(x)$ that yields a quartic double-well potential ${\rm Re}(V)=\beta( x^2-x_0^2)^2$ is shown in Fig.3(a). The energy spectrum of $\hat{H}=-\partial^2_x+\beta(x^2-x_0^2)^2+i \gamma x$ versus $\gamma$ shows a behavior similar to the Squire operator; in particular  above a threshold $\gamma_{PT}$  the spectrum ceases to be real, and pairs of real energies coalesce and become complex conjugates. Figure 3(b) shows the coalescence of the two lowest energy levels, corresponding to a symmetry breaking threshold $\gamma_{PT} \simeq 0.00065$. Above laser threshold, the onset of $\mathcal{PT}$ symmetry breaking corresponds to a transition from a stationary  to an oscillatory laser emission arising from the coexistence of two non-orthogonal competing transverse modes; see Fig.3(c). An elegant physical explanation of the phase transition in terms of Adler's theory of coupled oscillators  \cite{Adler1,Adler2} can be gained within a reduced two-mode model of the double well potential \cite{uffa}. The field $\psi(x,t)$ is approximated as $\psi(x,t) \simeq a_1(t) g_1(x)+a_2(t) g_2(x)$, where $g_{1,2}(x)$ are the ground-state wave functions localized in  each of the two wells of the potential. The evolution equations of amplitudes $a_{1,2}(t)$ are simply obtained from those of Refs.\cite{dimer1,dimer2} after Wick rotation, i.e. 
\begin{eqnarray}
\frac{da_1}{dt} & = & (\eta-i \sigma) a_1+\kappa a_2-\rho |a_1|^2 a_1 \\
\frac{da_2}{dt} & = & (\eta+i\sigma)a_2+\kappa a_1-\rho |a_2|^2 a_2
\end{eqnarray}
where $\kappa$ is the hopping rate between adjacent wells, $\eta=g_0-q$, $q$ is the decay rate in the absence of gain and coupling, $\sigma$ accounts for mirror tilting, and $\rho>0$ is the gain saturation parameter. Note that $2 \kappa$ measures the energy separation of the two levels in Fig.3(b) at $\gamma=0$,  $q$ is the mean values of such energies, and $\sigma$ is proportional to the non-Hermitian parameter $\gamma$ (mirror tilting). The optical power$P(t)$  emitted by the laser is proportional to $P \sim |a_1|^2+|a_2|^2$. The reduced model satisfactorily captures the spectral phase transition and the nonlinear dynamics above lasing threshold of the original Hamiltonian; see Figs.3(b,c). Interestingly, Eqs.(6) and (7) describe 
a canonical model of coupled oscillators near Hopf bifurcations, introduced for  chemical and biological dissipative  
systems \cite{vari} and studied in details by Aronson {\it et al.} \cite{Adler2}. The Wick-rotated $\mathcal{PT}$ symmetric dimer model basically describes two coupled laser oscillators \cite{Erneux}:  the two-humped  variable reflectivity mirror induces laser emission in two transverse regions, where the gain is maximum. Mode coupling is provided by evanescent field overlapping, whereas  mirror tilting controls the detuning of frequency emission of the two lasing regions. The transition observed in Fig.3(c) above the symmetry breaking is ultimately associated with breaking of oscillator phase locking. In fact, after setting $a_{1,2}=r_{1,2} \exp(i \phi_{1,2})$ and considering the symmetric 
solutions $r_1=r_2=r$, the coupled equations for the amplitude $r$ and relative phase $\phi=\phi_2-\phi_1$ of oscillators read (see Appendix C)
\begin{equation}
\frac{dr}{dt}=(\eta + \kappa \cos \phi)r-\rho r^3 \; ,\; \frac{d \phi}{dt}=2(\sigma - \kappa \sin \phi).
\end{equation}
The phase equation is the Adler's equation \cite{Adler1} that describes a transition from a phase-locked state $\phi={\rm asin}(\sigma / \kappa)$ for $\sigma < \kappa$ (i.e. $\gamma < \gamma_{PT})$, corresponding to a stationary amplitude $r$, to a phase drift state, associated with an oscillatory amplitude $r=r(t)$ for $\sigma> \kappa$. The signature of the spectral phase transition is thus universal phase locking-unlocking transition in Adler's theory of  coupled oscillators.

\section{Conclusions}
$\mathcal{PT}$ symmetry has provided a fruitful concept in different areas of physics. In this work the notion of $\mathcal{PT}$ symmetry has been extended by considering Wick rotation \cite{squire0}. In particular we have shown that light dynamics in optical cavities provides an accessible laboratory tool to explore $\mathcal{PT}$ phase transitions in Wick-rotated space. Wick rotation is known to be a useful tool in quantum physics to connect quantum mechanics and statistical mechanics \cite{Wick}.
 Here we suggest that Wick rotation can provide a fruitful  link between the physics of $\mathcal{PT}$ symmetric models \cite{Bender1} and pattern forming dissipative systems \cite{diss0,diss1}, thus broadening the class of physical systems where the notion of $\mathcal{PT}$ symmetry can be applied and experimentally explored.

\appendix

\section{Optical resonator model: mathematical aspects}
In this Appendix we derive the paraxial wave equation (3) given in the main text that describes transverse beam beam dynamics in the optical resonator of Fig.1. To this aim, let us assume that the laser oscillates on a single cavity axial mode and let us indicate by $E_m(X)$ the electric field amplitude at the transverse plane $\gamma$ and at the $m$-th round-trip in the cavity. The propagation of the field envelope in each round-trip can be readily obtained by considering the lensguide of Fig.A.4, which is obtained by unfolding the optical resonator starting from the plane $\gamma$. One can then write
\begin{equation}
E_{m+1}(X)=\hat{P} E_m(X)
\end{equation} 
where the round-trip operator $\hat{P}$ is given by
\begin{equation}
\hat{P}= \exp(g/2) r(X) \exp(g/2) \hat{K}.
\end{equation}
In the previous equation, $g/2$ is the single-pass saturated gain in the active medium (i.e. $g=\sigma_e d \Delta N$ where $\Delta N$ is the population inversion and $\sigma_e$ is the stimulated emission cross section of the laser transition), $r(X)=\sqrt{R(X)} \exp [i \Delta (X)]$ is the reflectivity of mirror 1, and $\hat{K}$ is the Huygens integral propagator that describes field propagation from plane $\gamma$ at $Z=0$ to plane $\beta$ in the lensguide of Fig.A.4. In writing Eq.(A.2), we neglected diffractive effects in the propagation across the gain medium, which is justified provided that the characteristic diffraction length of the resonator mode at plane $\gamma$ is much larger than $d$. The Huygens integral propagator $\hat{K}$ is given by \cite{Siegman}
\begin{equation}
\hat{K} f(X) = \sqrt{\frac{i}{\lambda B}} \int d\xi f(\xi)\exp \left[ -i \frac{\pi}{\lambda B} (DX^2+A \xi^2-2 \xi X) \right]
\end{equation}
where $ABCD$ is the paraxial generalized ray matrix from plane $\gamma$ to plane $\beta$. It can be readily calculated as the ordered product of the ABCD matrices of simple elements in the chain, and  reads explicitly
\begin{equation}
\left( 
\begin{array}{cc}
A & B \\
C & D
\end{array}
\right).= \left( 
\begin{array}{cc}
1 & 2 i \Theta f^2 \\
0 & 1
\end{array}
\right).
\end{equation}
In Eq.(A.4) we have set $\Theta= \lambda/ (\pi w_a^2)$, where $w_a$ is the size of the Gaussian aperture placed in the focal planes of the lenses. Note that the ABCD matrix is basically equivalent to the ray matrix describing free-space propagation over a 'complex' distance $2i\Theta$. With such in mind, it can be readily shown that the integral operator $\hat{K}$ can be cast in the following differential form
\begin{equation}
\hat{K}=\exp( \mathcal{D} \partial^2_X)
\end{equation}
where we have set
\begin{equation}
\mathcal{D}=\frac{2 \lambda \Theta f^2}{4 \pi}=2 \left( \frac{\lambda f}{2 \pi w_a} \right)^2.
\end{equation}
The operator $\hat{K}$ basically corresponds to a spectral filter, in Fourier space, of the near-field at plane $\gamma$. 
Equation (A.2) then yields
\begin{equation}
\hat{P}=\exp[g+ {\rm ln} \; r(X) ] \exp( \mathcal{D} \partial^2_X).
\end{equation}
Following Ref.\cite{App2}, we introduce a continuously evolving field amplitude 
$E(X,T)$, which depends on space $X$ and time $T$, such that $E(X,T=mT_R)$ is exactly equal to $E_m(X)$. The envelope $E(X,T)$ evolves in time according to the equation 
\begin{equation}
T_R \frac{\partial E}{\partial T}= \hat{\pi} E
\end{equation}
 where the operator $\hat{\pi}$ can be obtained from the relation $\exp( \hat{\pi}) =\hat{P}$. Using the  Baker-Campbell-Hausdorff formula, one has
 \begin{equation}
 \hat{\pi}=g+{\rm ln} [r(X)]+\mathcal{D} \frac{\partial^2}{\partial X^2}+\hat{e} 
 \end{equation}
 where we have set
 \begin{equation}
 \hat{e}=\frac{1}{2} \left[ \hat{P}_1,\hat{P}_2 \right]+\frac{1}{12} \left(  \left[ \hat{P}_1, \left[ \hat{P}_1, \hat{P}_2 \right] \right] -\left[ \hat{P}_2, \left[ \hat{P}_1, \hat{P}_2 \right] \right]  \right)+...
 \end{equation}
and $\hat{P}_1={\rm ln} \; r(X)$, $\hat{P}_2= \mathcal{D} \partial^2_X$. Assuming that the spectral filtering and the aspherical, variable-reflectivity mirror introduces small changes of the field $E$ in each round trip, at leading order the operator $\hat{e}$ in Eq.(S-9) provides a small correction and can be neglected \cite{App3,App4}, thus obtaining Eq.(3) given in the text. 
\\

\begin{figure*}
\includegraphics[width=16cm]{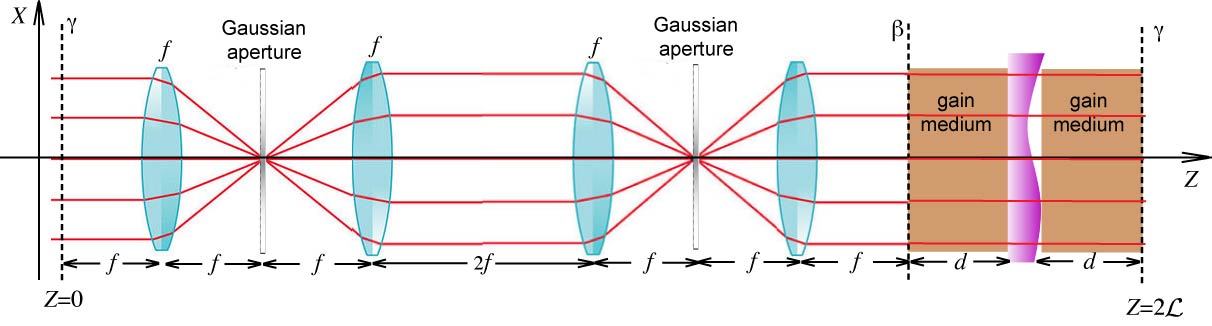}
\caption{(Color online) Lensguide associated with the resonator of Fig.1 in the main text. The round-trip operator $\hat{P}$ in Eq.(A.1) describes field propagation from plane $\gamma$ at $Z=0$ to the same plane at $Z=2 \mathcal{L}$, where $\mathcal{L}=4f+d$ is the cavity length. The Huygens integral operator $\hat{K}$ defined by Eq.(A.3) describes field propagation from plane $\gamma$ at $Z=0$ to plane $\beta$ at $Z=8f$. The transmission of the Gaussian aperture is $t(X)=\exp(-X^2 / w_a^2)$. Note that in the lensguide mirror 1 is replaced by a variable-transmission plate with transmission $t'(X)=r(X)=\sqrt{R(X)} \exp[i \Delta (X)]$, embedded in the two gain regions.}
\end{figure*}

 \section{Optical Squire model: Weakly nonlinear analysis.}  In this Appendix we briefly present a weakly nonlinear analysis to Eq.(1) for the optical Squire model close to laser threshold in the broken $\mathcal{PT}$ phase, where two transverse modes reach laser threshold simultaneously [see  Fig.2(a) for $\gamma>\gamma_{PT}$]  giving rise to mode competition. To this aim, let us assume $\gamma > \gamma_{PT}$ and let us indicate by $g_{0 \; th}$ the unsaturated gain parameter at the laser threshold, which is defined by the real part of the lowest energy $E$ in Fig.2(a).  Close to laser threshold, we look for a solution to Eq.(1) as a power series
\begin{equation}
\psi(x,t)= \epsilon \psi^{(1)}(x,t)+ \epsilon^3 \psi^{(3)}(x,t)+ ....
\end{equation}
where $\epsilon$ is a small parameter that measures the distance from the laser threshold, i.e. 
\begin{equation}
\epsilon^2 \equiv g_0-g_{0 \;th}.
\end{equation}
To avoid the appearance of secular growing terms in the asymptotic expansion (B.1), multiple time scales are introduced, namely $T_0=t$, $T_2= \epsilon^2 t$, .... Substitution of the Ansatz (B.1) into Eq.(1) and using the derivative rule $\partial_t=\partial_{T_0}+ \epsilon^2 \partial_{T_2}+...$ yields a hierarchy of equations for successive corrections to $\psi$. At leading order $\sim \epsilon$ one obtains
\begin{equation}
\partial_{T_0} \psi^{(1)}- \mathcal{L} \psi^{(1)}=0
\end{equation}
where $\mathcal{L}$ is the operator $-\hat{H}$ defined by Eq.(2) for $g_0=g_{0 \; th}$, i.e. $\mathcal{L}=\partial^2_x-V(x)+g_{0 \; th}$. The (non-decaying) solution to Eq.(B.3) can be written as a superposition of the two marginally-stable resonator modes $u_{1,2}(x)$ with amplitudes $a_{1,2}$ that depend on the slow time scale $T_2$, i.e.
\begin{equation}
\psi^{(1)}=a_1(T_2) u_1(X) \exp(i \Omega T_0) +a_2(T_2) u_2(X) \exp(-i \Omega T_0),
\end{equation}
where $\pm i \Omega$ are the complex-conjugate eigenvalues of $\mathcal{L}$ with eigenvectors $u_{1,2}(X)$, i.e.
\begin{equation}
\mathcal{L}u_{1,2}(X)=\pm i \Omega u_{1,2}(X).
\end{equation} 
Note that $2 \Omega$ corresponds to the frequency spacing between the two marginally-stable transverse modes. Note also that, owing to the $\mathcal{PT}$ symmetry of $\mathcal{L}$, one  has $u_2(x)=u_1^*(-x)$. At order $\sim \epsilon^3$ one obtains
\begin{equation}
\partial_{T_0} \psi^{(3)}- \mathcal{L} \psi^{(3)}=G^{(3)}
\end{equation}
where we have set
\begin{equation}
G^{(3)} \equiv -\partial_{T_2}  \psi^{(1)}+\psi^{(1)}-|\psi^{(1)}|^2 \psi^{(1)}.
\end{equation}
Substitution of Eq.(B.4) into Eq.(B.7) shows that the driving term $G^{(3)}$ entering in Eq.(B.6) can be written as 
\begin{equation}
G^{(3)}= A(x) \exp(i \Omega T_0)+ B(x) \exp(-i \Omega T_0) + ....
\end{equation} 
 where 
 \begin{eqnarray}
 A & \equiv & \left[  (-\partial_{T_2}+1)a_1-|a_1|^2a_1 |u_1|^2-2|a_2|^2 a_1 |u_2|^2 \right] u_1 \nonumber \\
 B & \equiv & \left[ (-\partial_{T_2}+1)a_2-|a_2|^2a_2 |u_2|^2 -2|a_1|^2 a_2 |u_1|^2 \right] u_2  \nonumber 
 \end{eqnarray}
 and where the dots stand for other terms oscillating like $\exp( \pm 3 i \Omega T_0)$.  To avoid the appearance of secularly growing terms in the solution to Eq.(B.6), that would prevent the validity of the asymptotic expansion (B.2), the following solvability conditions should be satisfied
 \begin{equation}
 \langle u_1^{\dag}(x) | A(x) \rangle =0 \; ,\;\;  \langle u_2^{\dag}(x) | B(x) \rangle =0 
 \end{equation}
 where $u_{1,2}^{\dag}(x)$ are the eigenvectors of the adjoint operator $\mathcal{L}^{\dag}$ corresponding to the eigenvalues $ \mp i \Omega$, and $\langle f |g \rangle = \int_{-\infty}^{\infty} dx f^*(x)g(x)$ is the ordinary (Hermitian) inner product.  Taking into account that $u_1^{\dag}(x)=u_1^*(x)$ and $u_2^{\dag}(x)=u_2^*(x)=u_1(-x)$, from the solvability conditions (B.9) one readily obtain
 \begin{eqnarray}
 \frac{da_1}{dT_2} & = & a_1-(\alpha |a_1|^2 +\beta |a_2|^2 ) a_1 \\
 \frac{da_2}{dT_2} & = & a_2-(\alpha^* |a_2|^2 +\beta^* |a_1|^2 ) a_2 
 \end{eqnarray}
 where we have set
 \begin{equation}
 \alpha = \frac{\int_{-\infty}^{\infty} dx u_1^2(x) |u_1(x)|^2}{\int_{-\infty}^{\infty} dx u_1^2(x)}\; , \;\;  \beta = 2\frac{\int_{-\infty}^{\infty} dx u_1^2(x) |u_1(-x)|^2}{\int_{-\infty}^{\infty} dx u_1^2(x)}
 \end{equation}
 If we stop the asymptotic analysis at this order and introduce the amplitudes $c_{1}(t)=\epsilon a_{1}(t)$, $c_{2}(t)=\epsilon a_{2}^{*}(t)$, one has 
 \begin{equation}
 \psi(x,t) \simeq c_1(t) u_1(x) \exp(i \Omega t) +c_2^*(t) u_2(x) \exp(-i \Omega t) + O(\epsilon^3)
 \end{equation}
 where the slowly-varying complex amplitudes $c_{1,2}(t)$ satisfy the nonlinear coupled equations
 \begin{eqnarray}
 \frac{dc_1}{dt} & = & (g_0-g_{0 \; th}) c_1-(\alpha |c_1|^2 +\beta |c_2|^2 ) c_1 \\
 \frac{dc_2}{dt} & = & (g_0-g_{0 \; th}) c_2-(\alpha |c_2|^2 +\beta |c_1|^2 ) c_2. 
 \end{eqnarray}
 Equations (B.14,B.15) admit of the following limit cycle solutions:\par
 (i) $c_1=\sqrt{(g_0-g_{0 \; th})/\alpha_R} \exp(i \delta t) $, $c_2=0$, with $\delta =-(\alpha_I / \alpha_R) (g_0-g_{0 \; th}$.\par
 (ii) $c_1=0$, $c_2=\sqrt{(g_0-g_{0 \; th})/\alpha_R} \exp(i \delta t) $, with $\delta =-(\alpha_I / \alpha_R) (g_0-g_{0 \; th}$.\par
 (iii) $c_1=c_2=\sqrt{(g_0-g_{0 \; th})/(\alpha_R+\beta_R)} \exp(i \delta t)$, with $\delta =-[(\alpha_I+\beta_I) /( \alpha_R+ \beta_R)] (g_0-g_{0 \; th}$.\\
 \\
In the above equations, $\alpha_{R,I}$ and $\beta_{R,I}$ denote the real and imaginary parts of the self- ($\alpha$) and cross- ($\beta$) saturation terms. Solutions (i), (ii) correspond to laser oscillation on a single transverse mode, either $u_1(x)$ or $u_2(x)$; this solution exists provided that $\alpha_R>0$ and is it stable for $\beta_R>\alpha_R$, i.e. if cross-gain saturation prevails over self-gain saturation. Solution (iii) corresponds to simultaneous oscillation of the  two transverse modes. Such a solution exists for $\alpha_R+\beta_R>0$ and it is stable for $\beta_R<\alpha_R$. The computation of the self- and cross-saturation coefficients $\beta_R$ and $\alpha_R$ can be done numerically using Eq.(B.12) once the mode profile $u_1(x)$ has been computed. For the case shown in Fig.2 [see the mode profiles in central and right panels of Fig.2(b)], one obtains $\alpha_R>\beta_R$ (namely $\alpha_R/ \beta_R \simeq 4.8$). This explains the result of Fig.2(e), where at $\gamma>\gamma_{PT}$ simultaneous oscillations of two transverse mode, leading to mode beating and oscillation of the output laser power, is observed. Note that power oscillation arises because the transverse modes are not orthogonal. In fact, using Eq.(B.13) the normalized output power $P(t)$ is readily calculated as 
\begin{eqnarray}
P(t) & =  & \int dx |\psi(x,t)|^2 \simeq  (|c_1|^2+|c_2|^2) \left( \int dx |u_1(x)|^2\right) \nonumber \\
& + & 2 {\rm Re} \left\{ c_1c_2^* \exp(2i \delta t) \int dx u_1(-x)u_1(x) dx \right\}
\end{eqnarray}
 The oscillating term, provided by the last term on the right hand side of Eq.(B.16), does not vanish whenever the inner product $\langle u_2(x)| u_1(x) \rangle=\langle u_1^*(-x)|u_1(x) \rangle \neq 0$, i.e. for a non-Hermitian operator $\mathcal{L}$.
 
 \section{Phase locking / phase drift transition in Wick-rotated $\mathcal{PT}$-symmetric dimer.} 
 After setting $a_{1,2}(t)= r_{1,2}(t) \exp [ i \phi_{1,2}(t) ] $, the nonlinear equations (6) and (7)  of coupled oscillators given in the text take the form
 \begin{eqnarray}
 \frac{dr_1}{dt} & = & \eta r_1+\kappa r_2 \cos \phi-\rho r_1^3 \\
 \frac{dr_2}{dt} & = & \eta r_2+\kappa r_1 \cos \phi -\rho r_2^3 \\
 \frac{d \phi}{dt} & = & 2 \sigma - \kappa \left( \frac{r_1}{r_2}+\frac{r_2}{r_1} \right)  \sin \phi 
 \end{eqnarray}
 where $\phi= \phi_2 - \phi_1$ is the relative phase of the two oscillators. An extended analysis of Eqs.(C.1-C.3) was presented by Aranson {\it et al.} in Ref.\cite{Adler2} in a more general framework. Here we just briefly review the main results of relevance for our analysis. Since asymmetric solutions $r_1 \neq r_2$ to Eqs.(C.1-C.3) are unstable \cite{Adler2}, we focus on the symmetric case $r_1=r_2=r$, so that one has
 \begin{eqnarray}
 \frac{dr}{dt} & = & \eta r+\kappa r \cos \phi-\rho r^3 \\
 \frac{d \phi}{dt} & = & 2 \sigma - 2 \kappa   \sin \phi.
 \end{eqnarray}
 In this case case the equation (C.5) for the relative phase $\phi$ of oscillators decouples from the amplitude equation (C.4) and has the form of Adler's equation, which is a gradient flow, i.e. $(d \phi /dt)=- \partial_{\phi} G$ with $G( \phi)=-2\sigma \phi-2\kappa \cos \phi$. This means that $(dG/dt)=-(d \phi / dt )^2 \leq 0$, i.e. in the dynamics the functional $G$ can not grow. For $\sigma < \kappa$, $G$ has relative minima and a stable global attractor of the dynamics is $\phi={\rm asin} ( \sigma / \kappa)$ with $-\pi/2 <  \phi  < \pi/2$   (apart from multiplies than $ 2 \pi$).  Conversely, for $\sigma > \kappa$ the function $G( \phi)$ does not show minima, the two oscillators can not be synchronized and the relative phase $\phi(t)$ drifts indefinitely in time. Once the solution $\phi(t)$ to the Adler equation has been determined, the amplitude equation (C.4) can be solved yielding the following general solution
 \begin{equation}
 r^2(t)=\frac{r^2(0) \exp \left[ 2 \eta t +2 \kappa  \int_0^t d \xi \cos \phi(\xi)  \right]}{1+ 2 \rho r^2(0) \int_0^t d \xi \exp \{  2\eta \xi + 2 \kappa  \int_0^ xi dq \cos \phi(q)  \}}.
 \end{equation}
 After an initial transient, the asymptotic behavior of Eq.(C.6) as $ t \rightarrow \infty$ can be readily calculated and reads
 \begin{equation}
 r^2(t) \sim \frac{1}{\rho} \; \frac{( \sigma^2+ \eta^2 - \kappa^2) \eta}{\sigma^2 + \eta^2- \kappa \eta \cos \phi(t) -\kappa \sigma \sin \phi(t)}.
 \end{equation}
 In the phase locking regime ($\sigma < \kappa$) the amplitude $r^2(t)$ settles down to the stationary value $r^2 = ( \eta / \rho) (\sigma^2+ \eta^2 -\kappa^2)/(\sigma^2+\eta^2-\kappa \eta)$, whereas in the phase drift regime ($\sigma > \kappa$) $r^2(t)$ is an  oscillatory function. The period $T_p$ of oscillation can be computed as the time interval needed to the phase $\phi(t)$ to drift from $\phi=0$ to $\phi= 2 \pi$, i.e.
 \begin{eqnarray}
 T_p & = & \int_{0}^{T_p} dt=\int_{0}^{2 \pi} \left( \frac{d \phi}{dt} \right)^{-1} d \phi \\
 & = &  \frac{1}{2} \int_{0}^{2 \pi} \frac{d \phi}{\sigma - \kappa \sin \phi} = \frac{\pi}{\sqrt{\sigma^2- \kappa^2}}. \nonumber 
 \end{eqnarray}

\end{document}